\documentclass[twocolumn,aps,showpacs,amsmath,superscriptaddress]{revtex4}
\usepackage{amssymb}
\usepackage{mathrsfs}
\usepackage{graphicx}
\usepackage{float,color}

\begin{document}

\title{Wigner crystal versus Fermionization for one-dimensional Hubbard models with and without long-range interactions}

\author{Zhihao Xu}
\email{xuzhihao@outlook.com}
\affiliation{Beijing National
Laboratory for Condensed Matter Physics, Institute of Physics,
Chinese Academy of Sciences, Beijing 100190, China}
\author{Linhu Li}
\affiliation{Beijing National
Laboratory for Condensed Matter Physics, Institute of Physics,
Chinese Academy of Sciences, Beijing 100190, China}
\author{Gao Xianlong}
\affiliation{Department of Physics, Zhejiang Normal University,
Jinhua, Zhejiang Province, 321004, China}
\author{Shu Chen}
\affiliation{Beijing National
Laboratory for Condensed Matter Physics, Institute of Physics,
Chinese Academy of Sciences, Beijing 100190, China}
\date{ \today}
\begin{abstract}
The ground state properties of Hubbard model with or without
long-range interactions in the regime with strongly repulsive
on-site interaction are investigated by means of the exact
diagonalization method. We show that the appearance of $N$-crests
in the density profile of a trapped N-fermion system is a natural
result of ``fermionization" between antiparallel-spin fermions in
the strongly repulsive limit and can not be taken as the only
signature of Wigner crystal phase, as the static structure factor
does not show any signature of crystallization. On the contrary,
both the density distribution and static structure factor of
Hubbard model with strong long-range interactions display clear
signature of Wigner crystal. Our results indicate the important
role of long-range interaction in the formation of Wigner crystal.
\end{abstract}

\pacs{ 03.75.Ss, 67.85.Lm,
71.10.Fd 
 }

\maketitle

\section{Introduction}
The Hubbard model has been widely used as a minimal model to
describe strongly interacting electrons in a solid and played a
particularly important role in understanding the physics related
to the quantum magnetism and high-temperature superconductivity.
Recent experiment progress in trapping the quantum gas in optical
lattice \cite{Greiner,Esslinger08,Koehl} has renewed the interest
in this basic model \cite{Jaksch,Hofstetter}. In comparison with
real materials in solid form, the fermionic quantum gas trapped in
optical lattice is a more ideal realization of the Hubbard model
\cite{Esslinger08}, which provides a laboratory to simulate and
study this basic quantum many-body system from different aspects.
Particularly current experiment techniques also prompt intensive
study of quantum many-body systems in reduced one-dimensional (1D)
systems, among which important benchmarks include experimental
realization of the Tonks-Girardeau (TG) gases
\cite{Paredes,Toshiya}, and the Mott transition in 1D tight tubes
\cite{Stoeferle}. Furthermore, the recent experimental realization
of two-component Fermi gases in a quasi-1D geometry with tunable
interaction strengths by Feshbach resonance \cite{Moritz05}
provides a unique possibility to experimentally study the 1D
Hubbard model even in the strong interaction limit. The good
tunability of optical lattices enables us to experimentally
elucidate the subtleties of 1D quantum many-body systems, for
example, the physical properties in the strongly interacting
limit. Due to recent experimental progress on chromium atoms
\cite{Thierry Lahaye,Stuhler} and heteronuclear polar molecule
produced by stimulated Raman adiabatic passage technique
\cite{K.-K. Ni,S. Ospelkaus}, dipolar atomic systems with strong
long-range dipole-dipole interaction (DDI) have also attracted
intensive studies \cite{Yi,Santos,Citro,Dalmonte,Zhihao
Xu,Deurezbacher}. In addition, schemes of creating Coulomb-like
interaction in cold atom systems have also been proposed \cite{D.
O'Dell,Maryvonne Chalony}.

In the presence of long-range repulsive interaction, an
interesting issue for the 1D fermionic system is the existence of
Wigner crystal phase \cite{Schulz}. For a 1D gas of electrons with
long-range Coulomb interaction which has been investigated by
Schulz in the scheme of bosonization \cite{Schulz}, the extremely
slow decay of density correlations at wave vector $4k_F$ has been
used as a signature of appearance of Wigner crystal. Here $k_F
=\pi n$ is the Fermi wave vector with $n$ the particle density.
After that, exploring Wigner crystal phase in the short-range
interacting fermion systems, for example the Hubbard model, has
also been carried out \cite{Eggert09,Eggert11,Gao
Xianlong07,Vieira}. By using bosonization and density matrix
renormalization group methods to study the one-dimensional finite
Hubbard chains with hard-wall boundaries, a crossover from $2k_F$-
to $4k_F$-density oscillations with the increasing of repulsion
has been found \cite{Eggert09,Vieira}, and sometimes the $4k_F$
density oscillations was taken as a signature of the emergence of
Wigner crystal. In the limit of infinitely strong interaction, the
on-site repulsion imposes an effective Pauli principle between
fermions with different spins. Consequently, the ground state in
the strong repulsion limit is degenerate for states with different
spin configuration \cite{Shiba,Chen,Girardeau07}, and thus the
ground state density distribution of an equally mixed trapped
system is identical to the fully polarized system with the
appearance of the doubling of the crests in density profiles. Such
a phenomenon, which is also refereed as fermionization, was
recently observed by the experiment in a two-particle system with
tunable interaction using two fermionic $^6$Li atoms \cite{Zurn}.

As the density distribution is not solely a convincing criteria of
Wigner crystal, the density-density correlation function plays an
important role in characterizing the Wigner crystal phase
\cite{Gao12}. In the present work, we shall study both the density
distribution and density correlation function of the Hubbard model
either without or with long-range interactions. For the Hubbard
model with only the short-range interaction, we find that the
density correlation function does not show any signature of
Wigner-crystal correlations in spite of the fact that the
ground-state density profile displays pronounced $4k_F$ Wigner
oscillations in the strong repulsion limit. To unveil the role of
the interaction range, we then study the Hubbard model with
additional long-range interactions and compare the calculated
results with those of the Hubbard model. We find that the system
with long-range interactions indeed exhibits quite different
behaviors from those of short-range Hubbard model. Particularly,
the clear evidence in static structure factor has indicated the
important role of the long-range interaction in the formation of
Wigner crystal.

The remain of the paper is organized as follows. In Sec.II, we
introduce our studied models and main conclusions based on the
calculation of models. In Sec.III, we present our calculation
results of density distributions and static structure factors in
subsection A and B for the Hubbard model and Hubbard model with
additional long-range interactions, respectively. A summary is
given in the last section.

\section{Model Hamiltonian}
The basic model describing the short-range interacting fermions in
a 1D lattice is the Hubbard model \cite{Lieb_Wu}
\begin{equation}
\label{eqn1}
H=-t\sum_{i\sigma}({\hat{c}^{\dag}_{i\sigma}\hat{c}_{i+1\sigma}+\mathrm{H.c.}})
+U\sum_{i}{\hat{n}_{i\uparrow} \hat{n}_{i\downarrow}},
\end{equation}
where $\hat{c}^{\dag}_{i\sigma}$($\hat{c}_{i}$) is the
creation(annihilation) operator of the fermion with the spin
$\sigma$, $\hat{n}_{i\uparrow}$ ($\hat{n}_{i\downarrow}$) is the
spin-up(down) particle number operator, $t$ is the hopping
strength which is set to the unit, and $U$ is the on-site interaction
strength.
The 1D Hubbard model can be realized by trapping a two-component Fermi
gas in a deep 1D optical lattice.

To study the long-range interaction, we consider the following
Hamiltonian
\begin{eqnarray}
\label{eqn2}
H&=&-t\sum_{i\sigma}({\hat{c}^{\dag}_{i\sigma}\hat{c}_{i+1\sigma}+\mathrm{H.c.}})
+U\sum_{i}{\hat{n}_{i\uparrow}\hat{n}_{i\downarrow}} \nonumber \\
& & +\frac{1}{2}V\sum_{i \neq j} {\frac{\hat{n}_i
\hat{n}_j}{|i-j|^{\alpha}}},
\end{eqnarray}
with $V$ the strength of the long-range interaction and $\alpha$
the decaying exponent.  For dipolar atoms confined in a 1D optical
lattices, the dipole-dipole interaction decays as $1/|i-j|^3$ with
the exponent $\alpha=3$ \cite{Zhihao Xu}, whereas $\alpha=1$
corresponds to the electron system with long-range Coulomb
interaction. In the present work, we consider the case with
repulsive on-site interaction $U>0$ and repulsive long-range
interaction $V>0$.

In the following section, we shall first calculate the density
distribution and the density-density correlation function for the
Hubbard model (\ref{eqn1}) with the on-site interaction strength
varying from weak to strong repulsion limit by exact
diagonalization method. In order to see the oscillation of the
density profile, we shall use open boundary condition (OBC) in the
calculation of density distribution, whereas periodic boundary
condition (PBC) shall be used in the calculation of structure
factor, where for the case of PBC, the distance of long-range
interaction should be min$(|i-j|,L-|i-j|)$.
In the limit of $U \rightarrow \infty$, the total
ground-state density distribution with different spin
configurations is identical to that of the polarized
noninteracting fermions, despite the fact that the spin-dependent
densities are different. This is consistent with the exact result
obtained via the exact construction of many-body wavefunction
\cite{Chen}. Our numerical diagonalization results also show that
the structure factor, which is the Fourier transformation of
density-density correlation function, in the infinite repulsion
limit also displays the same behavior as the fully polarized
fermions. These results suggest that the appearance of the
doubling of peaks of the equally mixing system (or 4$k_F$ Wigner
oscillation) in the infinite repulsion limit can not be taken as a
hallmark of the emergency of Wigner crystal. It is just a kind of
``fermionization" between fermions with different spins as a
result of the effective Pauli principle imposed by the infinite
repulsion.

The absence of Wigner crystal phase in the short-range Hubbard
model implies that the long-range interaction may play a very
important role in the formation of Wigner crystal phase. To unveil
its role, we then calculate the density distribution and the
density-density correlation function for the Hubbard model with
the additional long-range interaction described by Hamiltonian
(\ref{eqn2}). For conveniences, we shall focus our calculation to
the case with $\alpha=1$ for which the long-range interaction
decays more slowly than the dipole-dipole interaction. Keeping the
on-site interaction fixed, we find that the Wigner crystal phase
emerges with the increase of the long-range interaction strength
$V$. Finally, we also calculate the case with $\alpha=3$ which
exhibits similar behavior as the case of $\alpha=1$.

\section{Results and discussions}
\subsection{Fermi-Hubbard model}
\begin{figure}[tbp]
\includegraphics[height=8cm,width=9.5cm] {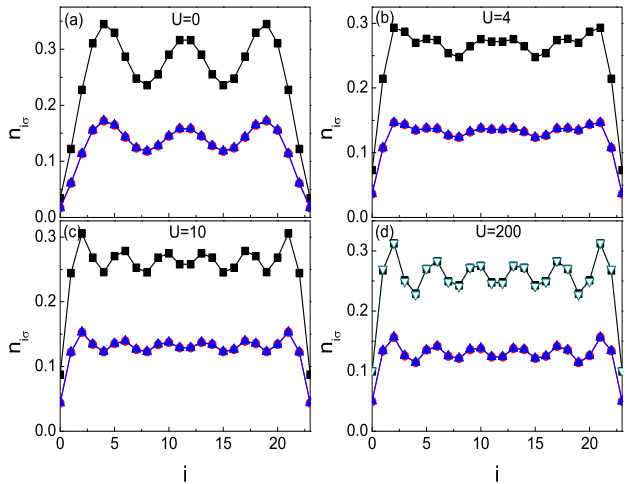}
\caption{(Color online) The ground-state density distribution of
the Fermi-Hubbard model for different $U$ with
$N_\uparrow=N_\downarrow=3$ confined in a lattice with $L=24$
under the open boundary condition. Here, solid square is for
$n_i=n_{i\uparrow}+n_{i\downarrow}$, solid circle is for
$n_{i\uparrow}$ and solid triangle is for $n_{i\downarrow}$. The
density distribution $n_i$ in the limit of $U \to \infty$ is shown
in (d) marked by hollow inverted triangles. }\label{Fig1}
\end{figure}

\begin{figure}[tbp]
\includegraphics[height=8cm,width=9.5cm] {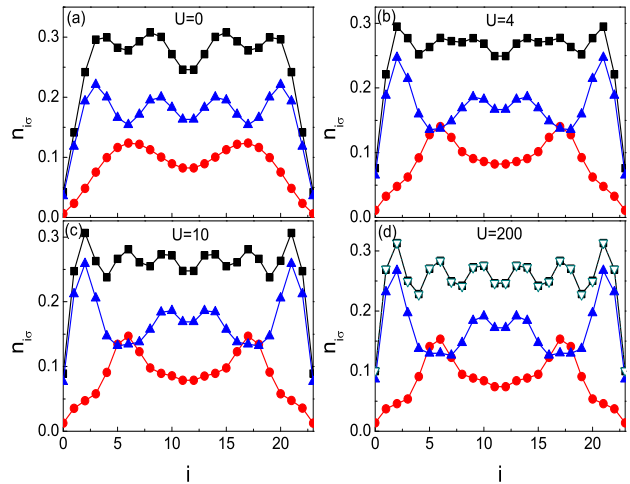}
\caption{(Color online) The density distribution of the Hubbard
model for different $U$ with $N_\uparrow=2,N_\downarrow=4$
confined in a lattice with $L=24$ under the open boundary
condition. Here, square is for $n_i$, circle is for
$n_{i\uparrow}$ and triangle is for $n_{i\downarrow}$. The density
$n_i$ for $U\to \infty$ is shown in (d) marked by hollow inverted
triangles. }\label{Fig2}
\end{figure}

We first calculate the ground state density distribution
$n_i=n_{i\uparrow}+n_{i\downarrow}$ of the Fermi-Hubbard model
(\ref{eqn1}) under the OBC, where $n_{i\sigma} = \langle
\hat{n}_{i\sigma} \rangle$ is the spin-dependent ground state
density distribution. In Fig.\ref{Fig1}, we show the density
profiles $n_i$ and $n_{i\sigma}$ versus different $U$ for the
balanced Fermi-Hubbard model (\ref{eqn1}) with $N_\uparrow =
N_\downarrow=3$ and the lattice size $L=24$.
For the balanced case, we always have
$n_{i\uparrow}=n_{i\downarrow}= n_i/2$. In the limit of $U
\rightarrow 0$, we have $n_{i\sigma}=\sum_{l}^{N_{\sigma}}
|\phi_{l}|^2$, where $\phi_{l}(i)=\sqrt{\frac{2}{L+1}}\sin(l
\frac{\pi}{L+1} i )$ is the $l$-th single particle wave function
of the open tight binding chain \cite{Busch}, which fulfills the
OBC of $\phi_{l}(0)=\phi_{l}(L+1)=0$. It is obvious that there are
three crests in the density distribution of $n_i$ for
$N_{\uparrow}=N_{\downarrow}=3$ in the weakly interacting limit.
As the interaction strength $U$ keeps increasing, the density
distribution displays clearly six crests for $U=10$. In the limit
of $U \rightarrow \infty$, we have $n_{i}=\sum_{l}^{N}
|\phi_{l}(i)|^2$ with $N=N_{\uparrow}+ N_{\uparrow}$ for different
spin configurations according to exact construction of many-body
wavefunctions \cite{Chen} due to the induced Pauli exclusion
principle between antiparallel spins. As shown in Fig.\ref{Fig1},
the total density profile for $U=200$ obtained by the exact
diagonaliztion method is completely overlapped  with the exact
result in the strongly interacting limit. For the imbalanced case
with $N_\uparrow=2$, $N_\downarrow=4$ and $L=24$, we show the
change of density profiles of the Hubbard model with the increase
of $U$ in Fig.\ref{Fig2}. Despite the fact that now we have
$n_{i\uparrow} \neq n_{i\downarrow}$, the total density
distribution also displays clearly six crests for $U=10$. Also,
the total density profile for $U=200$ coincides with the density
profile of the fully polarized system. Our numerical results
clearly show that the total density distributions of the
interacting fermion systems with different spin configurations are
identical in the limit of infinite repulsion \cite{Chen}.
\begin{figure}[tbp]
\includegraphics[height=8cm,width=9cm] {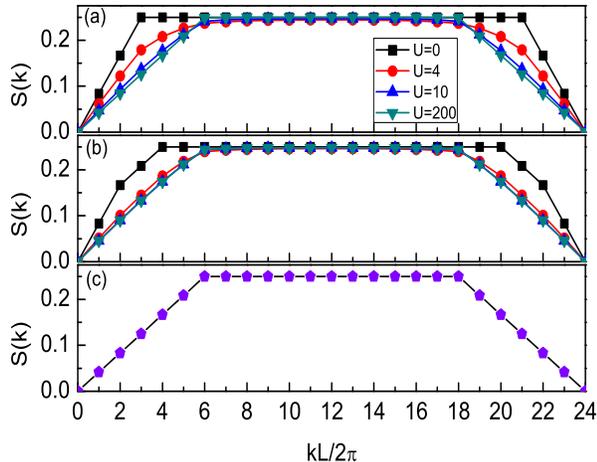}
\caption{(Color online) Static structure factor $S(k)$ vs
$kL/(2\pi)$ for different $U$ with the periodic boundary
condition. (a) is for $N_\uparrow=N_\downarrow=3,L=24$, (b) is the
case of $N_\uparrow=2,N_\downarrow=4,L=24$, and (c) is the case of
polarized free fermion system with $N=6$ and $L=24$.}\label{Fig3}
\end{figure}

Next we calculate the static structure factor, which is defined as
the Fourier transformation of the density-density correlation
function:
\begin{equation}
\label{eqn6} S(k)=\frac{1}{L}\sum_{i,j}{e^{ik(i-j)}[\langle
\hat{n}_i \hat{n}_j \rangle-\langle \hat{n}_i \rangle \langle
\hat{n}_j \rangle]},
\end{equation}
where $k=2m\pi/L$ with $m=0,1,...,L$. In general, the Wigner
crystal phase can be characterized by the enhancement of $4k_F$
peak in the the static structure factor. In Fig.\ref{Fig3}(a), we
show the change of the static structure factor with the increase
of interaction strength $U$ for the case of
$N_{\uparrow}=N_{\downarrow}=3$ and $L=24$ under the PBC. The
imbalanced case of $N_{\uparrow}=2$ and $N_{\downarrow}=4$ is
shown in Fig.\ref{Fig3}(b). For both cases, as $U$ is large enough
(for example, $U=200$), the distribution of the static structure
factor approaches to the distribution function of a fully
polarized $N$-particle free fermion system as shown in
Fig.\ref{Fig3}(c). Except of the drops at $kL/(2\pi)= N, L-N$, no
obvious peaks in the the static structure factor are detected.
Our numerical results clearly
indicate that only the appearance of $N$ peaks in the ground state
density profile of a finite trapped Hubbard system shown in
Ref.\cite{Eggert09,Eggert11,Gao Xianlong07,Vieira} can not be
taken as a mark of the emergence of Wigner crystal in the strongly
repulsive limit. However, as shown in next subsection, the Wigner
crystal phase appeared in the system with long-range interaction
exhibits an obvious 4$k_F$ peak in the static structure factor.

\subsection{Hubbard model with long-range interaction}
Now we study the Hamiltonian of the Hubbard model with long-range
interaction (\ref{eqn2}). For the case of $V=0$, the ground state
properties of the system almost does not change with the increase
of on-site interaction strength when the $U$ is large enough. In
order to explore the effect of the long-range interaction, we
shall keep $U$ fixed by taking $U=200$ and change $V$ in the
following calculation. For simplicity, we focus our discussion on
the case of $\alpha=1$, and finally discuss the case of $\alpha=3$
briefly.

\begin{figure}[tbp]
\includegraphics[height=6cm,width=9cm] {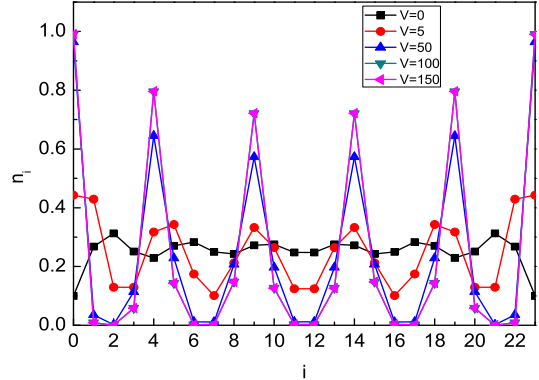}
\caption{(Color online) The density distribution of Hubbard model
with long-range Coulomb interaction for different $V$ with
$U=200$, $N_\uparrow=N_\downarrow=3$, and $L=24$ under the open
boundary condition. } \label{Fig4}
\end{figure}

In Fig.\ref{Fig4}, we show the density distribution for the
long-range Hubbard model with $\alpha=1$,
$N_\uparrow=N_\downarrow=3$ and $L=24$ under the OBC. At $V=0$,
six crests have already appeared for $U=200$. Nevertheless, with
the increase of $V$, the particles moves apart away from each
other as far as possible to minimizing the total energy. As shown
in the Fig.\ref{Fig4}, very sharp peaks have emerged for $V=50$
and the height of peaks increases with the increase of $V$. In
contrast to the case of $V=0$ with only  small oscillations, the
density profile of the system with a large $V$ resembles six
well-separated localized wave packets which distribute uniformly.
The density oscillation in Ref.
\cite{Eggert09,Eggert11,Gao Xianlong07,Vieira} corresponds to the
density distribution with $V=0$ shown in Fig.\ref{Fig4}, which
shows no sharp peaks and is very similar to the density
distribution of fully polarized fermions.

\begin{figure}[tbp]
\includegraphics[height=9cm,width=9cm] {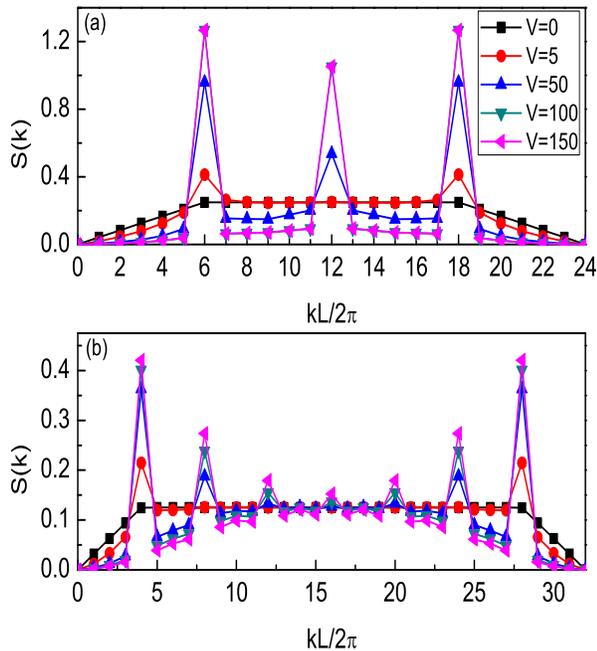}
\caption{(Color online) Static structure factor $S(k)$ vs
$kL/2\pi$ for systems with $U=200$ and different $V$ under the
periodic boundary condition. (a) is for $L=24$,
$N_\uparrow=N_\downarrow=3$. (b) is for $L=32$,
$N_\uparrow=1,N_\downarrow=3$. }\label{Fig5}
\end{figure}

\begin{figure}[tbp]
\includegraphics[height=6cm,width=9cm] {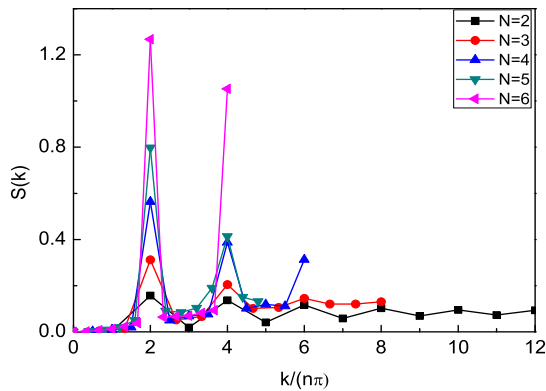}
\caption{(Color online) Static structure factor $S(k)$ vs
$k/(n\pi)$ ($n=N/L$) for systems with $U=200$ ,$V=150$, $L=24$ and
different $N$ under the periodic boundary condition. }\label{Fig6}
\end{figure}

To characterize the Wigner crystal phase, we further calculate the
static structure factor of the system with $L=24$, $N_\uparrow
=N_\downarrow=3$ under the PBC. As shown in Fig.\ref{Fig5}(a),
obvious peaks have already emerged at $kL/(2\pi)=N,L-N$ for $V=5$.
As $V$ keeps increasing, the height of peaks increases further and
new peak emerges at $kL/(2\pi)=2N$. In Fig.\ref{Fig5}(b), we
calculate the static structure factor for a system with $L=32$,
$N_\uparrow=1,N_\downarrow=3$ under the PBC. Similarly, as the
long-range interaction keeps increasing, a series of peaks emerge
at $kL/(2\pi)=mN$ with $m$ the integer.
We also demonstrate data for the static structure factor $S(k)$
versus $k/(n\pi)$ in Fig.\ref{Fig6} for systems with $L=24$,
$U=200$, $V=150$ and $N=2,3,4,5,6$, where $n=N/L$ is filling
factor of the system. As shown in Fig.\ref{Fig6}, systems with
different filling factors display similar behaviors in the regime
with strong long-range repulsive interaction, i.e., peaks emerging
at $k/(n\pi)=2m$. The Hubbard
systems studied in Ref. \cite{Eggert09,Eggert11,Gao
Xianlong07,Vieira} correspond to the case with $V=0$, for which no
sharp peaks emerge.

\begin{figure}[tbp]
\includegraphics[height=9cm,width=9.5cm] {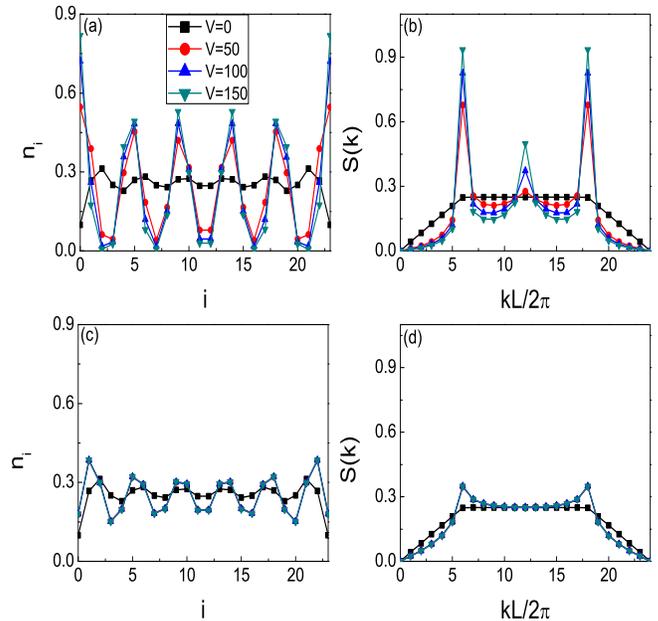}
\caption{(Color online) (a) Density profile of the Hubbard model
with dipole-dipole interaction under the open boundary condition;
(b)static structure factor of the Hubbard model with dipole-dipole
interaction under the periodic boundary condition. (c) density
profile of the extended Hubbard model with open boundary
condition; (d) static structure factor of the extended Hubbard
model with periodic boundary condition;  Here, $U=200$, $L=24$,
$N_\uparrow=N_\downarrow=3$. }\label{Fig7}
\end{figure}

Next we consider the Hubbard model with long-range dipole-dipole
interactions described by the Hamiltonian of Eq.\ref{eqn2} with
$\alpha=3$. In Fig.\ref{Fig7}(a), we plot the density profiles of
the Hubbard model with dipole-dipole interactions for a lattice
system with $L=24$, $N_\uparrow=N_\downarrow=3$ under OBC. The
static structure factors of the same system with PBC are plotted
in Fig.\ref{Fig7}(b). It is obvious that both the density profiles
and static structure factors display very similar behaviors as
these of the system with $\alpha=1$. A quantitative difference is
that the height of peaks in Fig.\ref{Fig7}(a) and (b) is lower
than that of the corresponding system with Coulomb interaction
shown in Fig.\ref{Fig4} and Fig.\ref{Fig5}(a), as the long-range
dipole-dipole interaction decays much faster than the Coulomb
interaction. Despite the minor differences, our calculated results
indicate the existence of the Wigner crystal phase for the Hubbard
model with strong dipole-dipole interactions.

As a comparison, we also study the extended Hubbard model
described by
\begin{equation}
\label{eqn7} H=-t\sum_{i\sigma}({\hat{c}^{\dag}_{i\sigma}
\hat{c}_{i+1\sigma}+\mathrm{H.c.}}) +U\sum_{i}{\hat{n}_{i\uparrow}
\hat{n}_{i\downarrow}}+V\sum_{i}{\hat{n}_i \hat{n}_{i+1}},
\end{equation}
where only the nearest neighbor short-range interaction is
considered.  Density profiles and static structure factors of the
extended Hubbard model with $L=24$, $N_\uparrow=N_\downarrow=3$
are shown in Fig.\ref{Fig7}(c) and (d), respectively. For $V=50$,
we observe that the oscillations in the density distribution of
the extended Hubbard model are enhanced, but the wave packets are
still overlapping, which is quite different from the density
profile of the dipolar system with well-separated localized
wave-packets. With further increasing the nearest neighbor
interaction, no obvious change is found. The static structure
factors also exhibit different behaviors from the long-range
interacting system. As shown in Fig.\ref{Fig7}(d), only small
peaks emerge at $kL/2(\pi)=N,L-N$ for $V=50$ and the height of
peaks does not change obviously with further increasing $V$. The
absence of peaks at $kL/(2\pi)=2N$ implies that the crystal phase
is not detectable. The different behaviors exhibited in systems
with the short-range and long-range interactions indicate the
important role of the long-range interaction in the formation of
Wigner crystal phase. By the way, the phase diagram for extended
Hubbard model in the half filling is studied widely\cite{Hirsch,
Nishimoto,Jeckelmann,Furusaki,Lin}.
The phase transition point between spin density wave(SDW) and charge
density wave(CDW) is $2V=U$ in the large $U,V$. Also we can find in
the large $U$, the quarter-filled system undergos a transition
from a Luttinger-liquid phase to a CDW with the increasing of $V$\cite{Zotos}.
At lower filling factor, a good approximation for large repulsive interaction
$U,V$, it is equivalent to the $N$ spinless fermions moving at $L-N$
lattice sites\cite{LinMingxi}.

To get a straightforward
understanding of the crystal phase of Hubbard model with
long-range interactions, we consider the atomic limit ($t=0$) of
the system with $L/N=1/n=\gamma$ under the PBC. In the atomic
limit, for the case of $V<U$ and $\gamma$ being an integer larger
than $1$, the ground states $|\psi_{0\mu}\rangle$ correspond to
the configurations with no more than two particles in the consecutive
$\gamma$ sites where $\mu=1,2\ldots\gamma\mathrm{C}_{N}^{N_{\uparrow\{\downarrow\}}}$
is for the different configurations of the degenerate ground states.
The degeneracy is derived from two parts, one is from the translation
invariance due to the PBC and the other comes from indistinguishable
spin configurations. The eigenenergy of
$H_0=U\sum_{i}{\hat{n}_{i\uparrow}
\hat{n}_{i\downarrow}}+\frac{1}{2}V\sum_{i \neq j}
{\hat{n}_i\hat{n}_j/(|i-j|^{\alpha})}$ is
$E_0=V/[(N/2)^{\alpha-1}\gamma^{\alpha}]+NV\sum_{\beta=1}^{N/2-1}1/(\beta\gamma)^{\alpha}$
for even $N$ and
$E_0=NV\sum_{\beta=1}^{(N-1)/2}1/(\beta\gamma)^{\alpha}$ for odd
$N$. To account for the first order perturbation of the hopping
term, we also consider the states $|\psi'_{0\mu m}\rangle$ which
are obtained from $|\psi_{0\mu}\rangle$ with one of the particles
hopping to the next sites and $m=1,2\ldots 2N$. The expected value
of $H_0$ in $|\psi'_{0\mu m}\rangle$ is $E_1=\langle\psi'_{0\mu
m}|H_0|\psi'_{0\nu l}\rangle=\delta_{\mu\nu}\delta_{m l}(E_0+E')$
where for $N=$ even
\begin{eqnarray}
\label{eqn2}
E'&=&\frac{V}{(\frac{N}{2}\gamma-1)^{\alpha}}-\frac{V}{(\frac{N}{2}\gamma)^{\alpha}} \nonumber \\
& & +V\Sigma_{\beta=1}^{N/2-1}\frac{1}{(\beta\gamma+1)^{\alpha}}+\frac{1}{(\beta\gamma-1)^{\alpha}}-\frac{2}{(\beta\gamma)^{\alpha}} \nonumber,
\end{eqnarray}
and for $N=$ odd
\begin{eqnarray}
\label{eqn2}
E'&=&V\Sigma_{\beta=1}^{(N-1)/2}\frac{1}{(\beta\gamma+1)^{\alpha}}+\frac{1}{(\beta\gamma-1)^{\alpha}}-\frac{2}{(\beta\gamma)^{\alpha}}
\nonumber.
\end{eqnarray}
The states of $\{|\psi_{0\mu}\rangle,|\psi'_{0\mu m}\rangle\}$
form a reduced space. The Hamiltonian $H=H_0+H'$ in such reduced
space is block diagonalization, and the dimension of each block is
$2N+1$, where
$H'=-t\sum_{i\sigma}(\hat{c}^{\dag}_{i\sigma}\hat{c}_{i+1\sigma}+\mathrm{H.c.})$
is the perturbation term. We compare the approximate results with
the one of exact diagonalization for
$N_{\uparrow}=N_{\downarrow}=3,L=24,t=1,U=200,V=150$. The relative
error of energy is about $0.08\%$ for $\alpha=1$.
The obvious peaks are observed at $k/(n\pi)=2m$ by calculating the
static structure factor using the perturbation wave functions,
characterizing the system in the Wigner crystal regime. While the
approximation does not work well with the same parameter for
$\alpha=3$ (the relative error of energy is about $27.03\%$), due to the
faster decay of the dipole-dipole interaction \cite{longrange}.
Here the strength of the interactions between particles with the
distance of $\gamma$ has the same order of magnitudes with the hopping amplitude,
so larger $U,V$ are needed for our perturbation approach.

\section{Summary}
In summary, we study the ground state properties of finite Hubbard
systems either with or without long-range interactions by using
the exact diagonalization method. We find that the appearance of
$N$-crests in the density profile of Hubbard model in the strongly
repulsive limit can not be taken as the only signature of Wigner
crystal phase, as it is induced by the effective Pauli principle
between antiparallel-spin fermions enforced by the infinite
repulsion. The absence of Wigner crystal phase is clearly verified
by the calculation of the static structure factor, which shows the
same behavior as the polarized free fermion system with no $4k_F$
peaks. If the long-range interaction is considered, the Hubbard
model with long-range interaction can form a perfect crystalline
phase in the regime with strong long-range repulsions. The
existence of the Wigner crystal phase can be well characterized by
its density distribution and the emergence of a series of peaks in
the static structure factor. Our study unveils the important roles
of the long-range interactions in the formation of the Wigner
crystal phase. In view of the rapid progress in manipulating the
dipolar atomic systems with tunable interactions, our results are
possible to be tested by the potential cold atom experiments.

\begin{acknowledgments}
This work has been supported by  National Program for Basic
Research of MOST, NSF of China under Grants No.11121063,
No.11174360 and No.10974234, and 973 grant.
\end{acknowledgments}


\begin{references}
\bibitem{longrange} For 1D fermionic models, the case of interactions with the $1/r^{\alpha}$-tails
has been studied by using bosonization method
\cite{Nomura,Kawakami}. Treating the long-range forward scattering
as a perturbation, they find the long-range forward scattering for
$\alpha>1$ is an irrelevant perturbation. We note that the above
conclusion is based on the perturbation analysis of the low-energy
effective Luttinger liquid theory with linear dispersion and it
does not exclude the possibility of the existence of Wigner
crystal in the limit of strong dipole-dipole interaction. Roughly
speaking, if $V/t \gg 2 \gamma^\alpha$, the long-range repulsion
tends to push the particles away from each other as far as
possible and the hopping term can not destroy the crystal phase.
\end{references}
\end{document}